\documentclass[aps,prl,reprint,superscriptaddress,amsmath,amssymb]{revtex4-1}

\usepackage{bm}
\usepackage{color}

\renewcommand{\vec}[1]{\bm{#1}}
\newcommand{\fermion}{\hat{\psi}}
\newcommand{\boson}{\hat{\phi}}

\newcommand{\schrodinger}{Schr\"{o}dinger }

\begin{document}

\title{Universal high-momentum asymptote and thermodynamic relations in a spinless Fermi gas with a resonant $p$-wave interaction}

\author{Shuhei M. Yoshida}
\affiliation{Department of Physics, University of Tokyo, 7-3-1 Hongo, Bunkyo-ku, Tokyo 113-0033, Japan}
\author{Masahito Ueda}
\affiliation{Department of Physics, University of Tokyo, 7-3-1 Hongo, Bunkyo-ku, Tokyo 113-0033, Japan}
\affiliation{RIKEN Center for Emergent Matter Science (CEMS), Wako, Saitama 351-0198, Japan}

\date{\today}

\begin{abstract}
We investigate universal relations in a spinless Fermi gas near a $p$-wave Feshbach resonance. We show that the momentum distribution $n_{\vec{k}}$ has an asymptote proportional to $k^{-2}$ with the proportionality constant--the $p$-wave contact-- scaling with the number of closed-channel molecules. We prove the adiabatic sweep theorem for a $p$-wave resonance which reveals the physical meaning of the $p$-wave contact in thermodynamics. In contrast to the unitary Fermi gas in which Tan's contact is universal, the $p$-wave contact depends on the short-range details of the interaction.
\end{abstract}


\maketitle


In the vicinity of an $s$-wave resonance of a short-range interaction, physical properties are known to become universal regardless of microscopic details of constituent particles and interparticle interactions. A gas of spin-$1/2$ fermions with a resonant $s$-wave interaction between the opposite spin components is called the unitary Fermi gas, and has attracted great interest in various fields of physics\cite{Zwerger2012}. Considerable efforts have been devoted to quantitatively determining universal properties of physical phenomena such as the universal equation of state. 
Another important aspect of the universality of the unitary Fermi gas is that there exist several \emph{universal relations} which connect the asymptotic short-range physics and thermodynamics of the system\cite{Tan2008a,Tan2008b,Tan2008c,Braaten2008a,Zhang2009,Combescot2009a,Werner2012}. They hold at any temperature and in whatever (normal or superfluid) phase. Here a single quantity, Tan's contact, encapsulates all needed information about a short-range correlation and its effect on thermodynamics. The universal relations have been examined both theoretically and experimentally\cite{Stewart2010,Kuhnle2010,Kuhnle2011,Hu2011,Sagi2012,Hoinka2013,VanHoucke2013,Hoinka2013}, and constitute the fundamental properties of the unitary Fermi gas.

Properties of systems with a resonant $p$-wave interaction have also been investigated. Such systems appear in nuclear and atomic physics. Among other things, ultracold atomic gases offer an interesting experimental platform on which a $p$-wave interaction can be tuned using a Feshbach resonance. A Feshbach resonance has been successfully observed and controlled in a gas of ${}^{6}\mathrm{Li}$ \cite{Zhang2004,Schunck2005} and that of ${}^{40}\mathrm{K}$\cite{Regal2003,Ticknor2004,Gunter2005}, and their scattering properties have been studied\cite{Chevy2005,Nakasuji2013}. Universal few-body physics near a $p$-wave resonance has also been discussed\cite{Jona-Lasinio2008,Nishida2012a,Braaten2012a,Nishida2013a}. The aim of this Letter is to address the universal properties of systems with a resonant $p$-wave interaction. We show that the momentum distribution has a high-momentum asymptote $n_{\vec{k}}\sim C_p k^{-2}$, and that the coefficient $C_p$, which we call a $p$-wave contact, is related to thermodynamics through a $p$-wave version of the adiabatic sweep theorem. We specifically discuss a spinless Fermi gas with a resonant $p$-wave interaction via a closed-channel molecular state. A crucial assumption in the following discussion is a separation of scales,
\begin{equation}
|v_m|^{1/3} > \lambda_T, k_F^{-1} \gg R_\mathrm{short},
\label{eq:scale-sep}
\end{equation}
where $v_m$ is a $p$-wave scattering volume which characterizes the effective strength of a $p$-wave interaction, $\lambda_T$ is the thermal de Broglie length, $k_F$ is the Fermi wave number, and $R_{\mathrm{short}}$ is a characteristic length scale of the short-range physics. In the case of neutral atoms, $R_{\mathrm{short}}$ is the largest among the van der Waals length, the inverse effective range, and a length scale set by a typical hyperfine splitting. The second inequality in (\ref{eq:scale-sep}) is naturally satisfied in ultracold atomic gases, while the first one is achieved using a Feshbach resonance.

Low-energy properties of the system 
are conveniently captured by a multi-channel model which explicitly incorporates 
the degrees of freedom of the closed-channel molecules relevant to the Feshbach resonance. 
However, if one naively takes the zero-range interaction limit, one would have negative probability states\cite{Pricoupenko2006,Jona-Lasinio2008,Hammer2010,Nishida2012a}. This is in contrast to the case of an $s$-wave resonance, in which the contact interaction model does not cause this problem. Therefore, we adopt the following Hamiltonian:
\begin{widetext}
\begin{equation} \begin{split}
\hat{H}
&= \int d^3\vec{r} \, \left\{
	\frac{\hbar^2}{2M}(\nabla \fermion)^2 
	+ \sum_{m=-1}^1 \left[
		\frac{\hbar^2}{4M} (\nabla\boson_m)^2 + \delta_m |\boson_m|^2
	\right]
\right\}  \\
&\quad + \sum_{m=-1}^1 \int d^3\vec{r} d^3\vec{r}' \, \left[
	\frac{\hbar^2 g}{2M} u_m( \vec{r}-\vec{r}' ) 
		\fermion^\dagger (\vec{r}) \fermion^\dagger(\vec{r}') 
		\boson_m\left(\frac{\vec{r}+\vec{r}'}{2}\right)
		+ \frac{\hbar^2 g^\ast}{2M} u_m^\ast( \vec{r}-\vec{r}' ) 
		\boson_m^\dagger\left(\frac{\vec{r}+\vec{r}'}{2}\right)
		\fermion(\vec{r}') \fermion (\vec{r}) 
	\right],
\label{eq:hamiltonian}
\end{split} \end{equation}
\end{widetext}
where the coupling function $u_m(\vec{r})$ is assumed to be finite and sufficiently short-ranged with the characteristic width typically comparable with the van der Waals length in the case of neutral atoms. Here, $\fermion$ is the fermionic field operator of the open-channel atoms with mass $M$ and $\boson_m$ is the bosonic field operator of the closed-channel molecules with the projected angular momentum $m$. The resonant interaction is characterized by the detuning parameter $\delta_m$, the coupling constant $g$ and $u_m(\vec{r})$. Due to the magnetic dipole-dipole interaction, the energy of a closed-channel molecule depends on $m$ in the presence of an external magnetic field, which is reflected in the $m$-dependence of $\delta_m$\cite{Ticknor2004}. The function $u_m(\vec{r})$ depends on the wave function of relative atomic motion in a molecule\cite{Gubbels2007}, and is normalized so that 
\begin{equation}
u_m(\vec{r}-\vec{r}') = \int \frac{d^3\vec{k}}{(2\pi)^3}
	e^{i\vec{k}\cdot(\vec{r}-\vec{r}')} k \chi(k^2) Y^1_m(\hat{\vec{k}}),
	\quad
\chi(0)=1.
\end{equation}
Since $u_m(\vec{r})$ extends over the van der Waals length 
$r_\mathrm{vdW}$, its Fourier transform has the following low-energy expansion:
\begin{equation}
\chi(k^2) \sim 1 + \chi_2 \frac{k^2}{\Lambda^2} ,
\end{equation}
where $\Lambda$ is the momentum scale of $O(R_\mathrm{short}^{-1})$ and $\chi_2$ a constant of $O(1)$.
Ignoring internal structures of the open-channel atoms and closed-channel molecules is justified by the latter inequality of Eq.~(\ref{eq:scale-sep}). 
The direct interaction between atoms in the open channel 
is also omitted. In reality, there is a long-range van der Waals potential for neutral 
atoms. The omission is justified if the inter-channel coupling is close to the Feshbach 
resonance and the open-channel interaction is far away from a shape resonance. 
These conditions are satisfied for ultracold gases 
of ${}^6\mathrm{Li}$ and ${}^{40}\mathrm{K}$\cite{Zhang2010}. 

The scattering amplitude of the two-body problem of the model can be obtained within an effective-range approximation as 
\begin{equation}
f(\vec{p},\vec{p}_0) 
= - \sum_{m=-1}^1 \frac{p^2 Y_m^1(\hat{\vec{p}}) Y_m^{1\ast}(\hat{\vec{p}}_0)}{1/v_m - k_e p^2 /2 + ip^3}.
\end{equation}
Here, $\vec{p}_0$ and $\vec{p}$ are the incoming and outgoing relative momenta, which satisfy $|\vec{p}|=|\vec{p}_0|\equiv p$, $v_m$ is a $p$-wave scattering volume, and $k_e$ is an effective range which has the dimension of momentum. Contrary to the usual collision via an isotropic potential, the scattering volume in the present case depends on the projected angular momentum $m$ of the relative motion of two colliding atoms because the symmetry is explicitly broken by the external magnetic field while the rotational symmetry about the $z$-axis remains. The bare parameters, $\delta_m$ and $g$, are written in terms of $v_m$ and $k_e$ as 
\begin{eqnarray} 
|g|^2 &=& \frac{32\pi^2}{|k_e|/2 - 16\pi^2 L_1 + 4\chi_2 /\Lambda^2 v_m}, \label{eq:g_squared} \\
\frac{M\delta_m}{\hbar^2} &=& \frac{-1/v_m + 16 \pi^2 L_3}{|k_e|/2 - 16 \pi^2 L_1}, \label{eq:detuning} 
\end{eqnarray}
where 
\begin{equation}
L_i \equiv \int \frac{d^3\vec{k}}{(2\pi)^3} k^{i-1} \chi(k^2)^2.
\end{equation}
In the vicinity of a Feshbach resonance, we can ignore the $1/v_m$-dependence of $g$. Then $g$ is solely determined by $k_e$, and therefore, the effective range stays nearly constant over the resonant region. On the other hand, the scattering volume in the $m$ channel varies as 
\begin{equation}
v_m = \frac{v_{m,\mathrm{bg}}\Delta B}{B-B_m}
\label{eq:Feshbach}
\end{equation}
as one sweeps the external magnetic field strength $B$ across a Feshbach resonance. Here, $B_m$ is the position of the Feshbach resonance, $\Delta B$ is its width, and $v_{m,\mathrm{bg}}$ is the background scattering volume which characterizes the residual interaction far away from the resonance. The quantity $v_{m,\mathrm{bg}}\Delta B$ in Eq.~(\ref{eq:Feshbach}) is related to the bare parameter by
\begin{equation}
v_{m,\mathrm{bg}}\Delta B = \frac{ |g|^2 }{ 32\pi^2 \Delta\mu_B }
\label{eq:FRwidth},
\end{equation}
where $\Delta\mu_B$ is the difference in the magnetic moment between the open and closed channels at the Feshbach resonance.

Now we turn to the many-body problem and derive a high-momentum asymptote of the momentum distribution. To achieve this goal, we consider the \schrodinger equation for the Hamiltonian (\ref{eq:hamiltonian}), and discuss its high-momentum behavior. Since the Hamiltonian conserves the total number $N$ of atoms in the open and closed channels to characterize an energy eigenstate, we need a set of wave functions $\Psi^{(N_o,N_c)}$ with $N=N_o+2N_c$ for $N_o$ open-channel atoms and $N_c$ closed-channel molecules. We impose the normalization condition
\begin{widetext}
\begin{equation}
\frac{1}{ N_o! N_c! } \int d^{3N_o}\vec{R} d^{3N_c}\vec{S} \sum_{ \{m_j=0,\pm 1 \} }
\left| \Psi^{(N_o,N_c)} ( \vec{R} , \vec{S}, \{ m_j \} ) \right|^2 = p(N_o, N_c)
\label{eq:normalization}
\end{equation}
on each $\Psi^{(N_o,N_c)}$, where $p(N_o,N_c)$ is the probability that the eigenstate has $N_o$ open-channel atoms and $N_c$ closed-channel molecules. Here, $\vec{R}$ and $\vec{S}$ collectively denote $\{ \vec{r}_i \}$ with $i=1,2,\dots,N_o$ and $\{ \vec{s}_j \}$ with $j=1,2,\dots,N_c$, respectively, and $m_j$ is the projected angular momentum of the $j$-th closed-channel molecule. The \schrodinger  equation is then written as
\begin{equation}
\hat{T} \Psi^{(N_o,N_c)} ( \vec{R} , \vec{S}, \{ m_j \} )
+ \hat{I}_{1} \left[ \Psi^{(N_o-2,N_c+1)} \right] ( \vec{R} , \vec{S}, \{ m_j \} )
+ \hat{I}_{2} \left[ \Psi^{(N_o+2,N_c-1)} \right] ( \vec{R} , \vec{S}, \{ m_j \} )
= E \Psi^{(N_o,N_c)} ( \vec{R} , \vec{S}, \{ m_j \} ) 
\label{eq:schrodinger-eq} ,
\end{equation}
where $\hat{T}$ is the kinetic energy plus the detuning,
and $\hat{I}_1$ and $\hat{I}_2$ correspond to the $\fermion^\dagger\fermion^\dagger\boson$ and $\boson^\dagger\fermion\fermion$ terms in the Hamiltonian, respectively.
To investigate the high-momentum behavior, we perform the Fourier transformation of Eq.~(\ref{eq:schrodinger-eq}) with respect to $\vec{r}_i$. 
Then the first and second terms on the left-hand side are multiplied by $k^2$ and $\tilde{u}_m(\mathbf{k})$, respectively, while the other terms do not acquire any additional factor of $k$ and become asymptotically negligible. We thus obtain
\begin{equation}
\frac{\hbar^2 k^2}{2M} \tilde{\Psi}^{(N_o,N_c)} ( \vec{k}, \tilde{\vec{R}}_i , \vec{S}, \{ m_j \} )
+ \frac{\hbar^2}{M} \sum_{j=1,j\neq i}^{N_o} \sum_{m=-1}^1 (-1)^{i+j+1} g \tilde{u}_m(\vec{k}) e^{i\vec{k}\cdot\vec{r}_j} \Psi^{(N_o-2,N_c+1)} ( \vec{R}'_{ij} , \vec{S}'_{ij}, \{ m_j \}' )
\simeq 0
\label{eq:asymp-eq}
\end{equation}
\end{widetext}
for large ${k}$. Here, $\tilde{\Psi}^{(N_o,N_c)}$ is the transformed wave function, and $\tilde{\vec{R}}_i$ is a collective notation of the coordinates of $N_o-1$ open-channel atoms other than the $i$-th atom. Due to the separation of the scales in Eq.~(\ref{eq:scale-sep}),  we can make the low-momentum expansion of the coupling function in Eq.~(\ref{eq:asymp-eq}) and obtain the asymptotic wave function. 
The momentum distribution of the open-channel atoms is then obtained by integrating the squared wave function with respect to $\tilde{\vec{R}}_i'$ and $\vec{S}$, summing over $ m_j $, and adding the contributions from all $\Psi^{(N_o,N_c)}$'s. This gives an asymptotic form,
\begin{equation}
n_{\vec{k}} 
\simeq \frac{C_p}{k^2} .
\label{eq:asymp-momentum-dist}
\end{equation}
We call the coefficient $C_p$ a $p$-wave contact in analogy with Tan's contact in the unitary Fermi gas. This scaling is implied in Ref.~\cite{Gurarie2007}, while their discussion is restricted to the superfluid phase and based on the saddle-point approximation. The explicit expression for the $p$-wave contact is obtained from Eq.~(\ref{eq:asymp-eq}) as
\begin{equation}
C_p = \frac{ |g|^2 }{ 4\pi } \langle \hat{N}_c \rangle 
= 8 \pi \sum_{m=-1}^1v_{m,\mathrm{bg}}\Delta B \langle \hat{N}_{c,m} \rangle,
\label{eq:contact}
\end{equation}
where we use Eqs.~(\ref{eq:normalization}) and (\ref{eq:FRwidth}), and $\hat{N}_c = \sum_{m=-1}^1 \hat{N}_{c,m}$. Equation~(\ref{eq:contact}) gives a physical meaning of the $p$-wave contact as the number of closed-channel molecules. 

The discussion so far is based on the \schrodinger equation and hence for a pure state. We can show the same relations as Eqs.~(\ref{eq:asymp-momentum-dist}) and (\ref{eq:contact}) by taking the average over the canonical ensemble. Therefore, these results hold not only for the ground state but also for thermal states.


An operational meaning of the $p$-wave contact is clarified by the adiabatic sweep theorem which indicates that the contact characterizes the response of the system to the interaction sweep by, for example, using a Feshbach resonance. Here, we show the theorem for the spinless $p$-wave system. First, from Eq.~(\ref{eq:detuning}), we obtain 
\begin{align}
\frac{\partial \hat{H}}{\partial (-1/v_m)} 
= \frac{ \hbar^2 |g|^2 }{ 32 \pi^2 M } \hat{N}_{c,m},
\end{align}
where $\hat{N}_{c,m}$ is the number operator of the closed-channel molecules with the projected angular momentum $m$.
Then, the Hellmann-Feynman theorem gives
\begin{equation}
\frac{\partial E}{\partial (-1/v_m)} 
= \left\langle \frac{\partial \hat{H}}{\partial (-1/v_m)} \right\rangle
= \frac{\hbar^2|g|^2}{32\pi^2 M} \left\langle \hat{N}_{c,m} \right\rangle
\label{eq:HFtheorem}
\end{equation}
for a pure state. Substituting this in Eq.~(\ref{eq:contact}), we finally obtain the adiabatic sweep theorem
\begin{equation}
C_p = \frac{8\pi M }{\hbar^2} \sum_{m=-1}^1
    \frac{ \partial E }{ \partial (-1/v_m) }
\end{equation}
for a pure state. 

The theorem can also be proven for a thermal state, or for a canonical ensemble in a similar manner as in Ref.~\cite{Werner2012}. Using the cyclic property of the trace, the isothermal derivative of the free energy by the scattering volume is given by
\begin{eqnarray}
\left( \frac{\partial F }{\partial (-1/v_m)} \right)_T
&=& Z^{-1} \mathrm{Tr} \left[
    e^{-\beta \hat{H}} \frac{\partial \hat{H}}{\partial (-1/v_m)}
\right] \nonumber \\
&=& \frac{ \hbar^2 |g|^2 }{ 32 \pi^2 M } \left\langle \hat{N}_{c,m} \right\rangle.
\label{eq:free-energy-derive}
\end{eqnarray}
Substituting Eq.~(\ref{eq:free-energy-derive}) in Eq.~(\ref{eq:contact}), we obtain the adiabatic sweep theorem for a thermal state,
\begin{eqnarray}
C_p = \frac{ 8\pi M }{\hbar^2} \sum_{m=-1}^1 \left( \frac{\partial F }{\partial (-1/v_m)} \right)_T \\
﻿= \frac{ 8\pi M }{\hbar^2} \sum_{m=-1}^1 \left( \frac{\partial E }{\partial (-1/v_m)} \right)_S,
\end{eqnarray}
where $F=E-TS$ and $T=\partial E/\partial S$ are used in obtaining the second equality.

Some comments are in order here. The power law in Eq.~(\ref{eq:asymp-momentum-dist}) seems too UV-singular to have a finite density. This is related to the fact that we cannot take the zero-range limit of a resonant $p$-wave interaction. The actual momentum distribution is cut off at the momentum scale $\Lambda\sim R_\mathrm{short}^{-1}$, and the resulting density is finite. The existence of the cutoff can be seen from the fact that the derivation is restricted to the region $|\vec{k}|\ll \Lambda$. Therefore, there is no actual ``UV divergence'' in the present treatment.

In connection with this point, we would like to remark the implication of the word, ``universality''. Our results are universal in that they are insensitive to the detailed form of the coupling function. However, this universality is weaker than that expected for the unitary Fermi gas, in which the thermodynamic function is believed to have a universal form regardless of the short-range detail of the system, and thus Tan's contact itself also has a universal behavior. In other words, physical quantities are well-defined in the limit of the infinite cutoff. However, this is not the case in the $p$-wave system. If we want to take the large cutoff limit, the $p$-wave contact converges to zero for the number of atoms to be finite. In a realistic system, we have a finite $p$-wave contact and the universal relations holds, but the value of the $p$-wave contact depends on short-range details such as $R_\mathrm{short}$. Therefore, our ``universality'' does not imply the universality of the $p$-wave contact itself.

In conclusion, we have derived a universal high-momentum asymptote in the momentum distribution, and identified its coefficient as the $p$-wave contact. We have also found that the $p$-wave contact has two physical meanings: the number of the closed-channel molecules and the derivative of the energy by the $p$-wave scattering volume. We point out that the universality with a resonant $p$-wave interaction is weaker than that with an $s$-wave interaction in that the value of the contact is sensitive to the short-range physics although the relations are insensitive to it.

This work was supported by KAKENHI Grant No. 26287088 from the Japan Society for the Promotion of Science, 
a Grant-in-Aid for Scientific Research on Innovation Areas ``Topological Quantum Phenomena'' (KAKENHI Grant No. 22103005),
the Photon Frontier Network Program from MEXT of Japan, and the Mitsubishi Foundation. S. M. Y. was supported by the Japan Society for the Promotion of Science through Program for Leading Graduate Schools (ALPS).

\bibliography{library}

\end{document}